# Cycloidal-spiral sampling for three-modal x-ray CT flyscans with two-dimensional phase sensitivity

G. Lioliou[*,1], O. Roche i Morgó[1], S. Marathe[2], K. Wanelik[2], S. Cipiccia[1], A. Olivo[1], C. K. Hagen[1]

[1]Department of Medical Physics and Biomedical Engineering, University College London, Malet Place, London WC1E 6BT, United Kingdom

[2]Diamond Light Source, Harwell Science and Innovation Campus, Fermi Avenue, Didcot OX11 0DE, UK

**Abstract**

We present a flyscan compatible acquisition scheme for three-modal X-Ray Computed Tomography (CT) with two-dimensional phase sensitivity. Our approach is demonstrated using a "beam tracking" setup, through which a sample's attenuation, phase (refraction) and scattering properties can be measured from a single frame, providing three complementary contrast channels. Up to now, such setups required the sample to be stepped at each rotation angle to sample signals at an adequate rate, to prevent resolution losses, anisotropic resolution, and under-sampling artefacts. However, the need for stepping necessitated a step-and-shoot implementation, which is affected by motors overheads and increases the total scan time. By contrast, our proposed scheme, by which continuous horizontal and vertical translations of the sample are integrated with its rotation (leading to a "cycloidal-spiral" trajectory), is fully compatible with continuous scanning (flyscans). This leads to greatly reduced scan times while largely preserving image quality and isotropic resolution.

**Introduction**

Three-modal X-Ray Computed Tomography (CT) delivers volumetric images in three (largely complementary) contrast channels: attenuation, phase (refraction), and small-angle scattering. Compared to conventional CT, which produces attenuation images only, the availability of the additional two channels is a key strength: phase (refraction) can show weakly attenuating details with a higher contrast-to-noise ratio (CNR) than the attenuation channel (at the same dose) [1], and scattering is sensitive to sample inhomogeneities below the imaging system's resolution [2], providing information that would otherwise be inaccessible. Owing to these advantages, three-modal CT is showing great promise for use in, e.g., biomedicine [3-5] and materials science [6].

---

[*] g.lioliou@ucl.ac.uk

Different technical solutions for three-modal CT exist [7-9]. Many of them use a modulator in the x-ray beam, typically placed upstream of the sample, to create a reference pattern that is then perturbed by the sample's attenuation, refraction, and scattering. Some techniques also employ some form of analyzer (a crystal, grating, or mask), placed downstream of the sample, to detect those disturbances. Generally, the three contrast channels are not measured directly but must be retrieved from the raw images [10] before further processing, e.g., CT reconstruction, can take place. When an analyzer is used, the retrieval requires a minimum of three images (frames) to be acquired with the analyzer in different positions, to obtain three pixelwise equations that can be solved for the three unknowns (attenuation, refraction, scattering) [11-15]. In CT, the need to acquire $n \geq 3$ frames per rotation angle leads to extended scan times, as the necessary repositioning of the analyzer enforces a step-and-shoot acquisition which incurs overheads. In other words, flyscans, characterised by a continuous sample rotation, cannot be implemented. This is problematic because: 1) the overheads impose a scan time bottleneck that cannot be overcome even when reducing the exposure time, and 2) it limits the throughput capabilities of three-modal CT as well as options for time-resolved imaging.

More recently, methods have emerged that allow retrieving three-modal x-ray images from a single frame [16-25]. In these, the analyzer is replaced by a high-resolution detector, featuring pixels small enough to resolve the reference pattern created by the modulator, and therefore allowing to extract the changes caused by the sample's attenuation, refraction, and scattering without acquiring and processing multiple frames. One of these methods is "beam tracking", an x-ray implementation of the Hartmann mask [26], in which the modulator is a mask with alternating absorbing and transmitting septa, generating beamlets with a spatial separation sufficient to make their mutual interference negligible. In this scenario, the sample's attenuation, refraction, and scattering lead to an intensity reduction, shift, and broadening of the beamlets, respectively. If a mask with long, narrow slits is used [22,23], beam tracking is sensitive to phase effects (refraction, scattering) in the direction orthogonal to the slits, while two-dimensional phase sensitivity is achieved with a mask with square or round apertures ("holes") [24,25].

A drawback of beam-tracking is that the parts of the sample covered by its absorbing areas cannot contribute to the image. Previously, this was handled by applying a "dithering" procedure, by which the sample is scanned in steps smaller than the mask period until all its parts have been illuminated by the x-ray beam, a frame is acquired at each step, and all frames are combined into an up-sampled image. When using a mask with slits, a one-directional scan is needed (in the direction orthogonal to the slits); when using a mask with holes, the sample must be scanned along two directions to obtain a fully sampled image. Although effective, this process is not compatible with flyscans when applied in CT, as dithering must be applied at each angle forcing corresponding interruptions in the sample's rotation. On the other hand, if no dithering is applied (i.e., by only rotating the sample in the beamlets), data are under-sampled, which has a detrimental effect on overall image quality and spatial resolution, with the latter then being driven by the mask period rather than by the apertures.

We have previously proposed cycloidal CT as an alternative to dithering for a setup where the modulator is a mask with long, narrow slits [27,28]. In a cycloidal scan, the sample is translated in the direction orthogonal to the rotation axis, continuously and simultaneously with being rotated. This was shown to provide a resolution comparable to dithering, while also being flyscan compatible, as it involves the acquisition of only a single frame per angle (contrary to dithering, where multiple frames are acquired). So far, cycloidal CT had only been used with "edge illumination" [29], a phase imaging technique that also uses a mask as the modulator but requires a geometrically magnified replica of the mask as an analyser, placed in front of a detector with pixels much larger than the beamlets. The need to acquire three frames per angle, each with a different offset between the masks, to retrieve attenuation, refraction and scattering signals [30] has so far prevented us from performing three-modal flyscans, despite the cycloidal approach's flyscan compatibility. Rather, the flyscan CT data we have acquired up to this point show a single "hybrid" contrast channel (a merge of attenuation and refraction), enabled by a retrieval method that relies on approximations and assumptions on the sample [31].

With this article, we build on the introduction of cycloidal CT by proposing a "cycloidal-spiral" acquisition scheme, by which the sample is translated along *two directions*, continuously and simultaneously with its rotation. To facilitate the acquisition of three-modal CT images through flyscans, we have combined this scheme with beam tracking, rather than edge illumination. This is implemented using a mask with holes, which provides the added benefit of two-dimensional phase sensitivity. In the following, we report on the optimization of the approach on a phantom (425-500 μm diameter polyethylene spheres in a 3 mm diameter plastic straw) and provide a proof-of-concept demonstration of its flyscan compatibility on a complex biological sample (a piglet esophagus).

**Results and discussion**
**Beam tracking and sampling schemes.** The beam tracking setup is shown in Figure 1. It consists of the x-ray source, the mask (placed upstream to the sample), the sample, and the detector (see **Methods** section). As mentioned, such a setup would normally be used in conjunction with dithering, although simple rotation-only scans are also possible. The former provides fully sampled data but is slow as the sample must be scanned in sub-period steps, $\Delta x_{opt}$ along the horizontal and $\Delta y_{opt}$ along the vertical direction, at each angle (the steps typically match the size of the apertures); the latter are fast and flyscan compatible, but image quality is compromised, and resolution driven by the mask period. Fig. 2(a) and 2(b) show the sampling grids for the two approaches, depicted for one mask period (but repeating itself for adjacent periods in either direction). The filled squares indicate the sub-period "slots" of the 3D dataset for which data are acquired (i.e., where a beamlet passes through the sample), while the empty squares indicate points where no data is available (i.e., corresponding to sample areas covered by the absorbing regions of the mask).

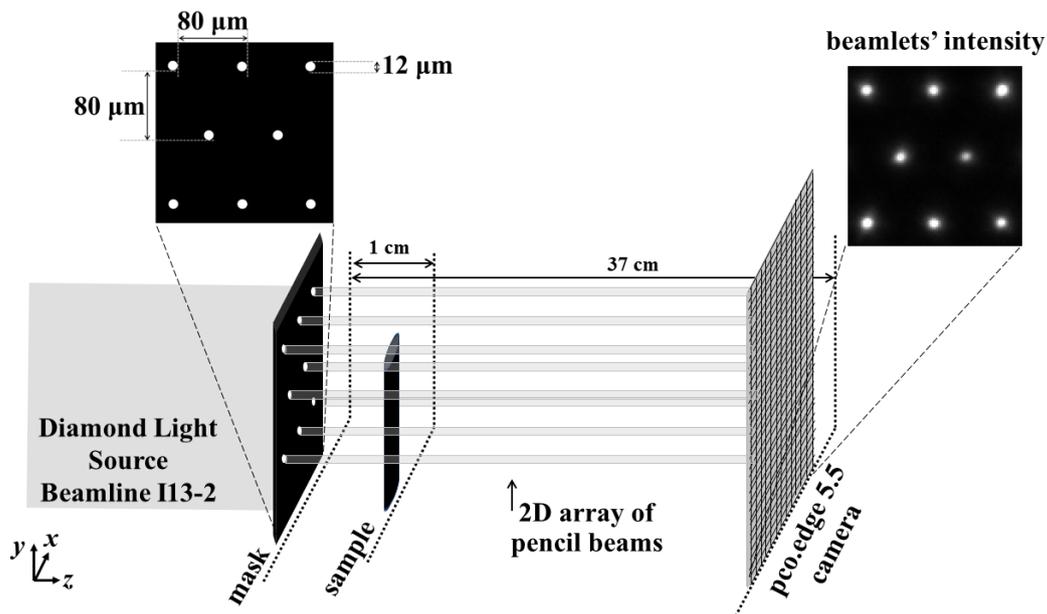

**Fig. 1.** Schematic (not to scale) of the experimental setup.

The proposed cycloidal-spiral scan shares similarities with a rotation-only scan in that only a single frame is acquired per angle. However, with each rotation, the sample is displaced by horizontal and vertical distances of *dx* and *dy*, hence a beamlet probes the sample in a different sub-period position, which creates an interlaced sampling pattern like the one shown in Fig. 2(c). The schematic shows an example case where $dx = \Delta x_{opt}$ and $dy = 4\Delta y_{opt}$. More generally, *dx* and *dy* can be flexibly chosen, so long as they are fractions of the mask period.

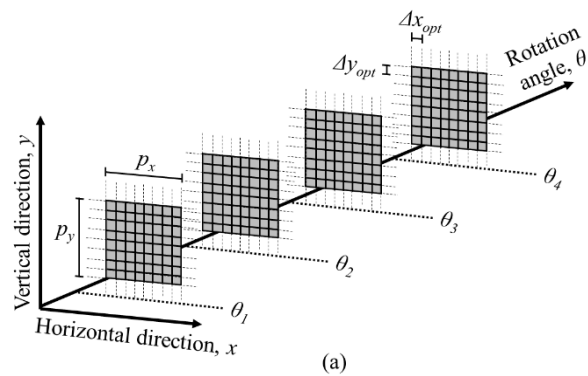

(a)

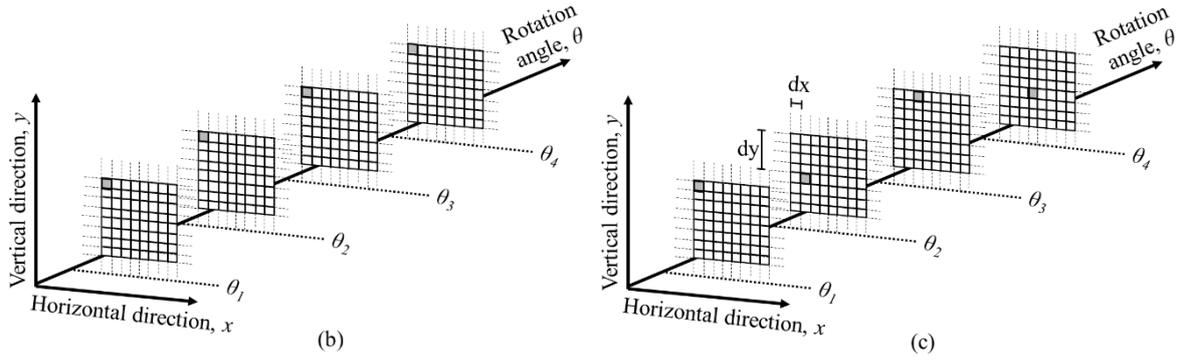

**Fig. 2.** Sampling grids (shown only for a single mask period in the horizontal, $p_x$, and vertical, $p_y$, direction) for a dithered scan (a), a rotation-only scan (b), and a cycloidal-spiral scan (c). In the former case, the sample is scanned in horizontal and vertical steps of $\Delta x_{opt} = p_x/8$ and $\Delta y_{opt} = p_y/8$, respectively, at each rotation angle. In the latter case, the sample is displaced by horizontal and vertical (sub-period) distances of $dx = \Delta x_{opt}$ and $dy = 4\Delta y_{opt}$, simultaneously with being rotated to the next angle.

**Cycloidal-spiral optimization.** To determine which values of sample displacement per rotation angle ($dx$ and $dy$) would deliver the best results, we initially acquired a dithered dataset of the sphere phantom (scan duration of 7.5 h) and sub-sampled it according to cycloidal-spiral patterns by assuming different combinations of $dx$ and $dy$, as per Table I; see **Methods** section. Note that the parameters $dx = dy = 0$ represent a rotation-only scan, which is therefore inherently part of the investigation.

The outcomes (reconstructed axial, sagittal, and coronal slices, see **Methods** section) of the sub-sampling are shown in the Supplementary Material; the images were compared in terms of their visual quality as well as their spatial resolution along the three axes of the reconstructed volume. Since the sphere phantom consisted of a weakly attenuating, homogeneous material (polyethylene), which meant there was little contrast in the attenuation and scattering channels (images not shown), the analysis was conducted on the phase channel. The resolution estimates (see **Methods** section) are listed in Table I. Among the different combinations of $dx$ and $dy$, the best results were obtained for $dx = dy = 20$ μm. In this case, resolution was estimated to be 28 μm, 40 μm, 31 μm along $x$, $y$, $z$, respectively, which, given there is a degree of uncertainty in these estimates, may be considered in line with our isotropy target. For $dx = dy = 0$ μm (equivalent to a rotation-only scan), the estimates were 47 μm, 101 μm, and 49 μm, which not only is a worse outcome, but it is also much less isotropic. For reference, resolution in the fully sampled images was estimated at 11 μm, 12 μm and 13 μm, respectively. Fig. 3 (a-l) shows the results for $dx = dy = 20$ μm, $dx = dy = 0$ μm, and the fully sampled (dithered) case.

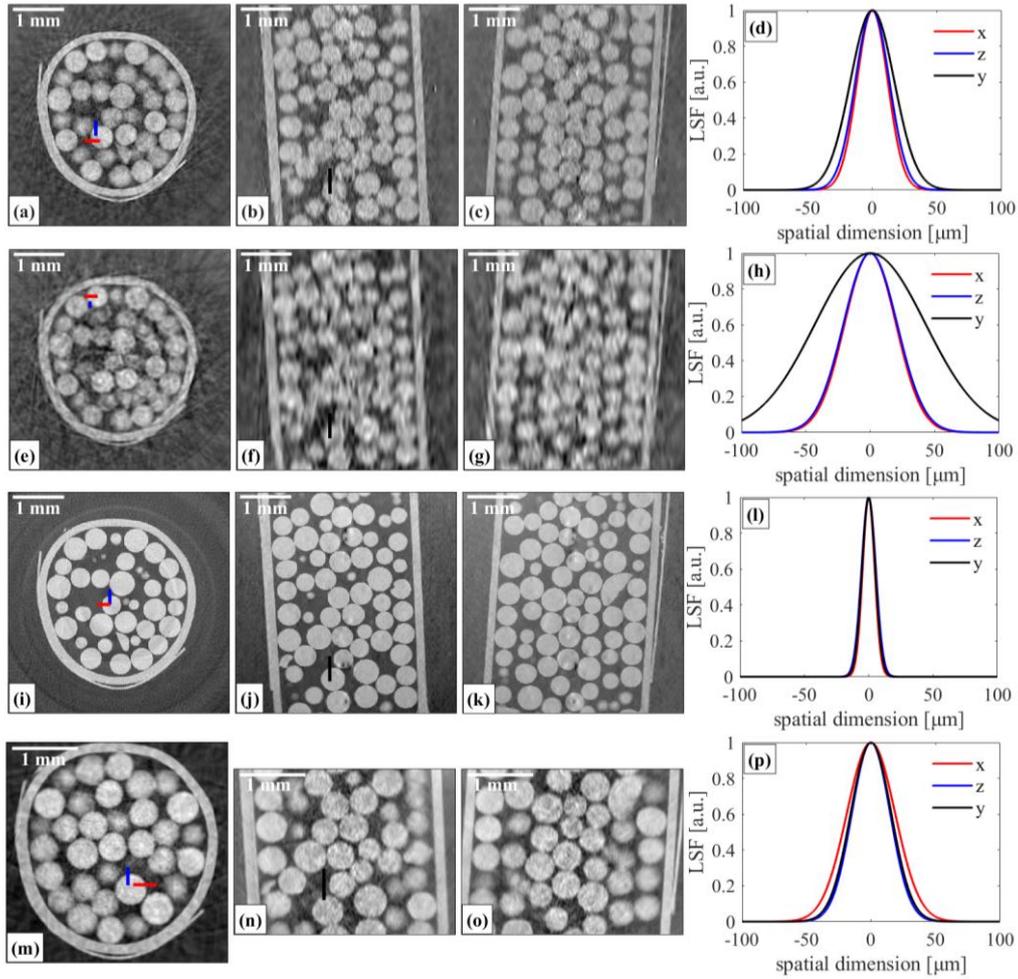

**Fig. 3.** Reconstructed axial (left), sagittal (central left), and coronal (central right) planes of the sphere phantom, and the corresponding line spread functions (right) extracted from the sphere edges indicated with corresponding colours along the *x*, *y* and *z* axes. Results are shown for a cycloidal-spiral sub-sampling with $dx = dy = 20$ μm (a-d); a rotation-only sub-sampling, i.e., $dx = dy = 0$ μm (e-h); the fully sampled (dithered) case (scan duration of 7.5 h) (i-l); a cycloidal-spiral flyscan performed with $dx = dy = 20$ μm (scan duration of 200 s) (m-p).

**Table I**. Spatial resolution estimates along the *x*, *y*, and *z* direction extracted from the phase images of the sphere phantom, obtained through sub-sampling the dithered dataset using the values of *dx* and *dy* given in the left most columns.

| *dx* (μm) | *dy* (μm) | Spatial resolution (μm) | | |
|---|---|---|---|---|
| | | *x* | *y* | *z* |
| 0 | 0 | 47 | 101 | 49 |
| 10 | 10 | 40 | 73 | 37 |
| 10 | 20 | 45 | 50 | 40 |
| 10 | 40 | 40 | 72 | 39 |
| 20 | 10 | 45 | 33 | 39 |
| 20 | 20 | 28 | 40 | 31 |
| 20 | 40 | 47 | 48 | 53 |
| 30 | 30 | 34 | 63 | 36 |
| 40 | 10 | 56 | 33 | 50 |
| 40 | 20 | 61 | 46 | 51 |
| 40 | 40 | 55 | 58 | 49 |

**Three-modal flyscans.** Informed by the above results, we then performed true three-modal flyscans with *dx* = *dy* = 20 μm (see **Methods** section); the scan duration was 200 s.

Flyscan images (phase channel only) of the polyethylene spheres are shown in Fig. 3(m-o). The overall quality of these images appears to be in line with their step-and-shoot counterparts, with no obvious artefacts arising from the continuous scanning. Resolution was estimated to be 44 μm, 37 μm, and 35 μm in *x*, *y* and *z*, respectively, implying that (near) isotropy was preserved; this is shown in Fig. 3(p).

Using the same parameters and procedures, we then scanned a more complex sample: a decellularised piglet esophagus (a so-called "scaffold"), which had been prepared in the context of tissue engineering research (Savvidis et al. [32]), to demonstrate the ability to extract three-modal images from a single flyscan (Fig. 4). Also in this case, spatial resolution appears to be isotropic in all three contrast channels. Of note is the greater CNR in the phase images (Fig. 4b), relative to the attenuation and scattering ones.

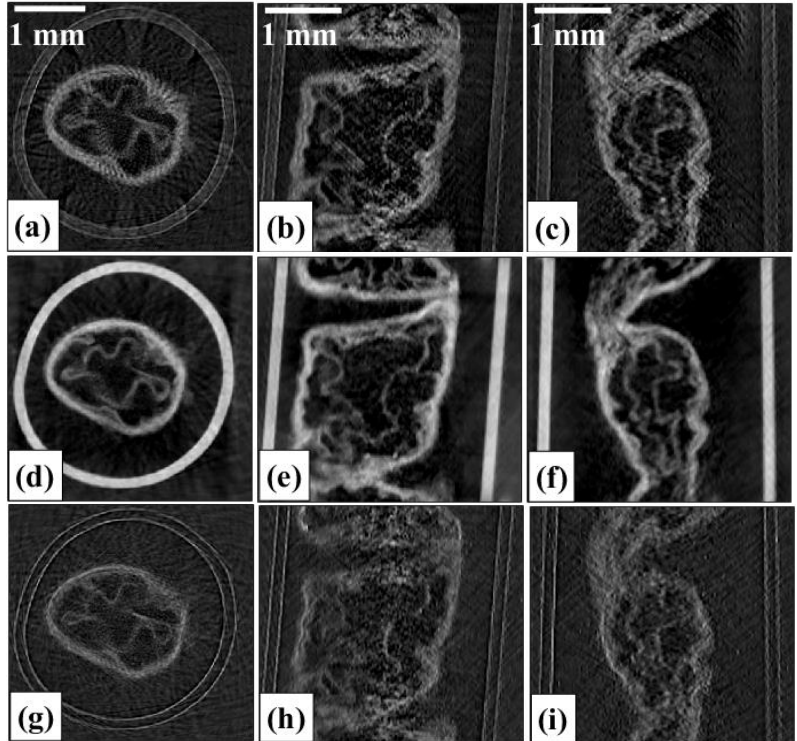

**Fig. 4.** Reconstructed (a) – (c) attenuation, (d) – (f) phase, and (g) – (i) scattering, (a), (d), (g) axial, (b), (e), (h) sagittal, and (c), (f), (i) coronal slices.

**Conclusions**

In summary, we presented three-modal flyscan x-ray CT enabled by combining a beam-tracking setup with two-dimensional phase sensitivity with a cycloidal-spiral acquisition scheme. Previously, similar setups required the implementation of step-and-shoot acquisitions, at least if high, isotropic resolutions were to be achieved. This, however, incurred significant overheads and imposed a scan time bottleneck that cannot be overcome by decreasing the exposure time. By contrast, in a flyscan, the overall scan time is determined by exposure time alone, as dead times are typically negligible.

The key outcome of the work is the reduction in scan time: we demonstrated that the cycloidal-spiral flyscan can be completed in a 135x faster time than a fully sampled (dithered) scan (scan duration of 200 s for the cycloidal-spiral flyscan c.f. 7.5 h for the fully sampled scan), while maintaining the ability to reconstruct three contrast channels with (near) isotropic resolution. Although the flyscan results had a lower resolution than their dithered counterpart (44 μm, 37 μm, and 35 μm in *x*, *y* and *z*, respectively for the cycloidal-spiral flyscan c.f. 11 μm, 12 μm and 13 μm in *x*, *y* and *z*, respectively for the dithered step and shoot scan), it must be considered that the former were reconstructed from 1/64[th] of the frames (and therefore overall amount of data) of the latter, hence some degree of degradation is to be expected. At the same time, the cycloidal-spiral approach was shown to be advantageous over a rotation-only scan. While both schemes can be implemented as flyscans (thus there is no difference in the overall scan time), the simple addition of sample motion within the plane parallel to the modulator during the rotation increases resolution and enables resolution isotropy. This is likely because the added translation "spreads" the acquired data more evenly across the 3D sinogram, thus improving the de facto sampling, while in a rotation-only scan data are densely packed along $\theta$ but spread coarsely along $x$ and $y$.

For this proof-of-concept demonstration, the cycloidal-spiral and rotation-only datasets were completed using simple interpolation, but related work, e.g., on using machine learning [33], suggests that more advanced methods could ultimately lead to a better performance.

Looking forward, we believe that the cycloidal-spiral approach could open new applications for three-modal CT that are currently unattainable due to constraints on scan time. The flyscans presented in this paper took 200 s, but it is reasonable to assume that this can be brought down to tens of seconds or below by reducing the exposure time per frame, at least at synchrotrons where there is an abundance of x-ray flux. Scan time would then be at a level where some degree of time-resolved imaging becomes feasible, a goal that will we seek to attain in the future. Furthermore, the cycloidal-spiral approach has positive implications for three-modal CT performed with lab sources (to which beam tracking can be adapted to [23]), where the much lower flux available exacerbates the problem of extended scan times, with scans currently lasting hours. By implementing flyscans, the acquisition of three-modal CT images with a few tens of µm resolution may become possible within minutes. Indeed, the translation of our method from the synchrotron to a laboratory environment will be another key focus of our future work.

**Methods**

**Experimental setup.** The experimental setup is shown in **Fig. 1**. All scans were performed at the Diamond Light Source, Beamline I13-2, using a pink beam with a spectrum in the 8 keV to 30 keV range and centered around 27 keV. The sample was placed at 210 m from the source [34], and the mask 1 cm upstream of the sample. The mask consists of a 200 ± 20 µm thick Au layer on a 1 mm thick graphite substrate (Microworks, Germany). It has 12 µm circular apertures with a period of 80 µm in both directions; every other aperture row had an offset of half the horizontal period (i.e., 40 µm, see Fig. 1). The experimental setup further consisted of a pco.edge 5.5 camera coupled to a scintillator-objective combination leading to an effective pixel size of 2.6 x 2.6 µm$^2$. The distance between the sample and detector was 37 cm.

**Data acquisition.** The dithered dataset consisted of 400 projections taken by rotating the sample in steps of 0.9 degrees over 360 degrees. At each angle, the rotation was interrupted, and the sample scanned in steps of $\Delta x_{opt} = \Delta y_{opt} = 10$ µm across one mask period, $p$, in each direction (step and shoot scan). This led to the acquisition of 64 frames per angle, with an exposure time of 0.5 s each. In total, 25,600 frames were acquired during this scan, which took 7.5 hours (including overheads arising from the scan's step-and-shoot nature).

The cycloidal-spiral flyscans also consisted of 400 projections (i.e., $d\theta = 0.9$ degrees), but now only a single frame was acquired per angle, as the sample was translated in the *x-y* plane while being rotated. The exposure time per frame also remained unchanged (0.5 s). Since in a flyscan all three sample motors (rotation, horizontal translation, vertical translation) are moved continuously and simultaneously, their speeds needed to be carefully tuned to the exposure time to achieve the desired sampling steps. The continuous acquisition allowed for negligible overheads to be incurred, meaning that the entire scan was completed in 200 s, which is 135x faster than the dithered scan.

**Data analysis.** Before the dithered dataset was sub-sampled, to replicate rotation-only and cycloidal-spiral scans, attenuation, two-dimensional refraction (along *x* and *y*), and two-dimensional scattering (along *x* and *y*) signals were retrieved from the raw frames by tracking each beamlet's profile and quantifying the changes induced by the sample. Specifically, attenuation was obtained from the beamlets' intensity reduction, and refraction from their horizontal and vertical displacements (through subpixel image registration based on cross correlation [35]), respectively. The scattering signal, here defined as the variance of the sample's scattering function along the horizontal, $\sigma_x^2$, and vertical, $\sigma_y^2$, direction was obtained from the variance of the beamlets' second moments [25].

The sub-sampling, which was applied individually to the retrieved contrast channels, involved discarding all but one of the 64 frames acquired per angle. At angle *j*, we only kept the frame, *i*, with the horizontal (*h*) and vertical (*v*) indices:

$$i_h = \left(\frac{((j-1)dx) \bmod p}{dx_{opt}}\right) + 1 \qquad [1a]$$

$$i_v = \left(\frac{((j-1)dy) \bmod p}{dy_{opt}}\right) + 1, \qquad [1b]$$

where *j*=1,…,400, as this created the desired cycloidal-spiral patterns. Consequently, the sub-sampled sinograms were 3D matrices with "gaps", containing only 1/64th ≈ 1.56% of the entries of their fully sampled counterparts. The missing 63/64th ≈ 98.44% of entries were filled through interpolation, for which two schemes were investigated. The first one consisted of applying 1D cubic spline interpolation along the *y* direction of the sub-sampled sinograms, followed by 2D cubic interpolation in the *x-θ* planes. The second one consisted of applying 2D cubic interpolation in the *y-θ* planes, followed by 2D cubic interpolation in the *x-θ* planes. Both schemes were applied for the interpolation of the missing entries in all sub-sampled 3D data sets; by comparing the results, it was found that the first scheme provides a greater image quality in the reconstructed axial planes, while the second scheme performs better in the coronal and sagittal planes. Thus, the axial planes presented in this article resulted from interpolation with the first scheme, while the coronal and sagittal planes resulted from use of the second. Following interpolation, the sub-sampled and fully sampled sinograms contained the same number of entries.

In all cases, CT reconstruction was performed with filtered back projection (FBP), applied on a slice-by-slice basis. Prior to reconstruction, the refraction images in *x* and *y* were converted into integrated phase images through a Fourier space method [36]. For simplicity, only the scattering signal along the horizontal direction, $\sigma_x^2$, was reconstructed and presented below.

In principle, the attenuation, refraction, and scattering signals extracted from the flyscan frames can be placed into their sub-period "slots" in the 3D sinogram using the indices calculated via Eqs. 1a and 1b. However, this approach cannot account for any discrepancy between the actual and nominal speeds of the translation stages, which would cause an inaccurate allocation. Moreover, in our specific case, we had to implement "back-and-forth" translations along both directions to prevent the 3 mm wide sample from exiting the 6.7 mm x 5.6 mm field of view. This involved moving the sample by 2 mm along *x* and *y*, reversing the translators' direction, moving the sample by 2 mm in the opposite directions, reversing translators reversed again and so forth, until the sample's 360-degree rotation had been completed. Owing to this, we found that there was indeed a discrepancy between actual and nominal sample trajectories, which became more pronounced with every turn of the sample, due to the motors' deceleration and acceleration. To take this into account, we calculated the indices as:

$$k_h = \frac{(CP_h - IP_h) \bmod p}{\Delta x_{opt}} \qquad [2a]$$

$$k_v = \frac{(CP_v - IP_v) \bmod p}{\Delta y_{opt}}, \qquad [2b]$$

where CP and IP refer to the sample's current and initial position in mm, i.e., its position at the time of acquiring a particular frame and at the beginning of the scan, respectively. It should be noted, however, that CP cannot be inferred from the acquired frames directly, as the horizontal and vertical translations of the sample are entangled with its rotation. Motor encoder readout synchronized to the detector acquisition can output CP with a high-level of accuracy but, if this is not available, the sample's translation must be tracked or measured explicitly. In our case, we repeated the flyscan on a reference sample (a tungsten sphere, diameter 300 μm) without the modulator and without the rotation, but with otherwise identical scan parameters. The horizontal and vertical positions of the tungsten sphere during that scan were tracked by applying subpixel image registration [35] to each pair of consecutive frames, and subsequently used as CP in Eqs. 2a and 2b. Once the attenuation, refraction and scattering signals extracted from the flyscan data had been correctly placed into the 3D sinograms, interpolation and CT reconstruction were applied as described above.

Resolution estimates were obtained by fitting error functions to the edges (indicated in the corresponding Supplementary Material figure for each case) along *x*, *y*, and *z* of spheres, computing their derivatives to obtain line spread functions (LSF), and extracting their resulting full width at half maxima (FWHM).

**Acknowledgments**

This work was in part supported by the Wellcome/EPSRC Centre for Interventional and Surgical Sciences (WEISS) (203145/Z/16/Z). Additional funding by EPSRC [EP/T005408/1] is acknowledged. C. K. Hagen is supported by the Royal Academy of Engineering, under the Research Fellowship scheme. A. Olivo is supported by the Royal Academy of Engineering, under the Chair in Emerging Technologies scheme (CiET1819/2/78). The authors gratefully acknowledge the provision of beamtime MG28574 on ID13-2 at the Diamond Light Source, and the assistance of the beamline scientists. The authors thank the ICH stem cell and regenerative medicine (SCRM) group, University College London, for the oesophageal sample preparation.

**Data availability**

The data that support the findings of this study are available from the corresponding author upon reasonable request.

**Author contributions**

C.K.H., A.O., and S.C. conceived the study. C.K.H., A.O., S.C., G.L., O.R.iM., and S.M. designed and performed the experiment. S.C. and K.W. wrote the data acquisition scripts. C.K.H. and G.L. prepared the figures for the manuscript. C.K.H., A.O., and G.L. interpreted the data and wrote the manuscript with the input from all authors.

**Competing interests**

The authors declare no competing interests.

**Supplementary material**

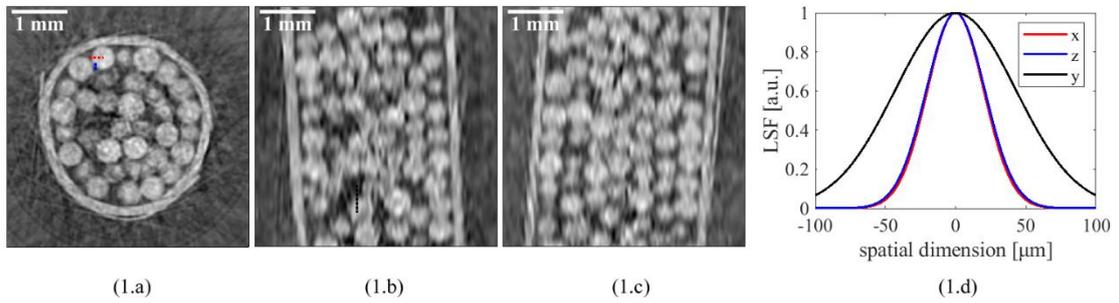

(1.a) (1.b) (1.c) (1.d)

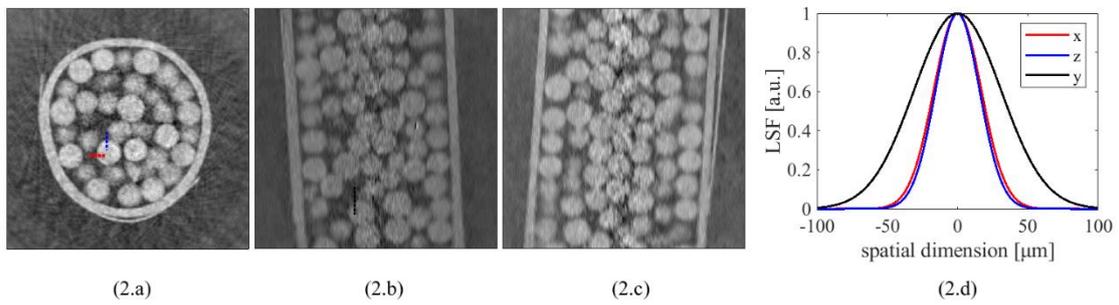

(2.a) (2.b) (2.c) (2.d)

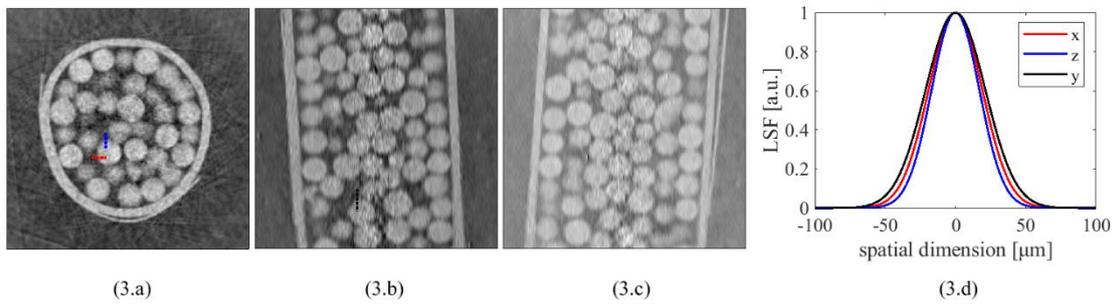

(3.a) (3.b) (3.c) (3.d)

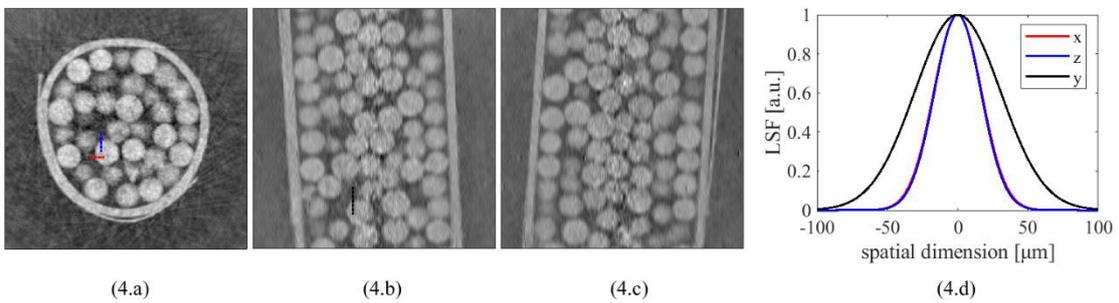

(4.a) (4.b) (4.c) (4.d)

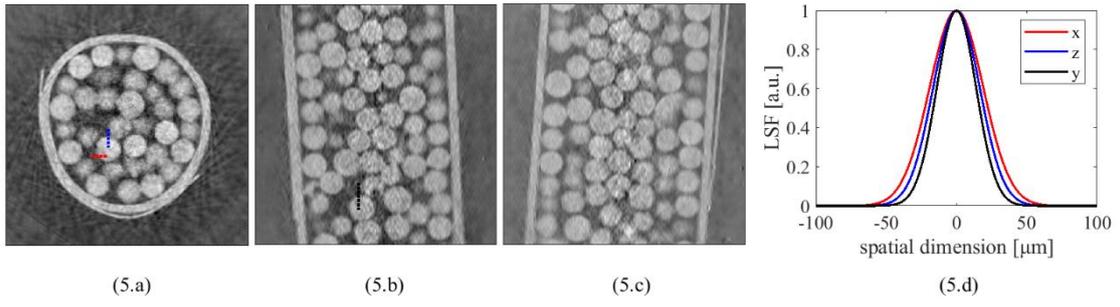

(5.a)　　　　　(5.b)　　　　　(5.c)　　　　　(5.d)

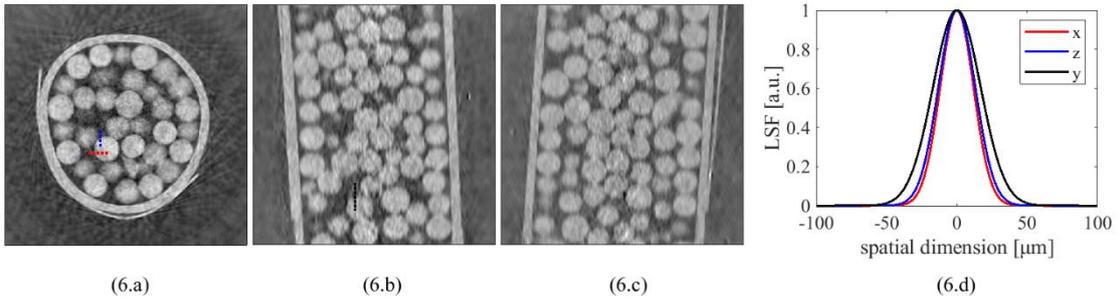

(6.a)　　　　　(6.b)　　　　　(6.c)　　　　　(6.d)

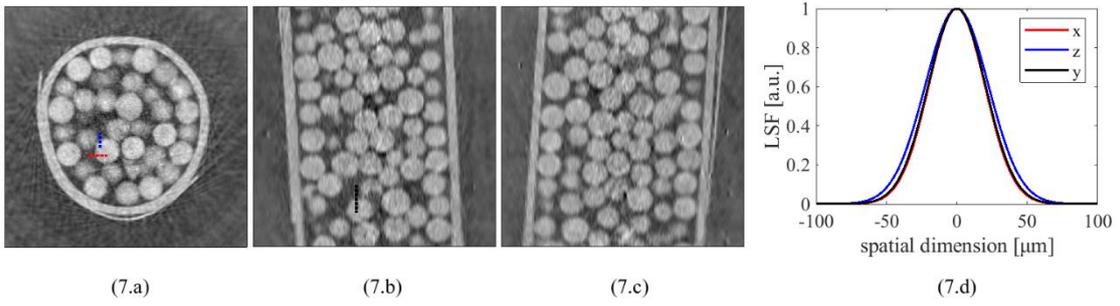

(7.a)　　　　　(7.b)　　　　　(7.c)　　　　　(7.d)

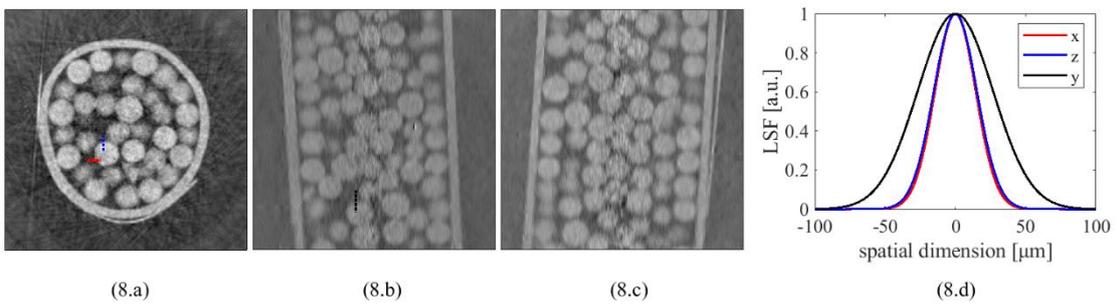

(8.a)　　　　　(8.b)　　　　　(8.c)　　　　　(8.d)

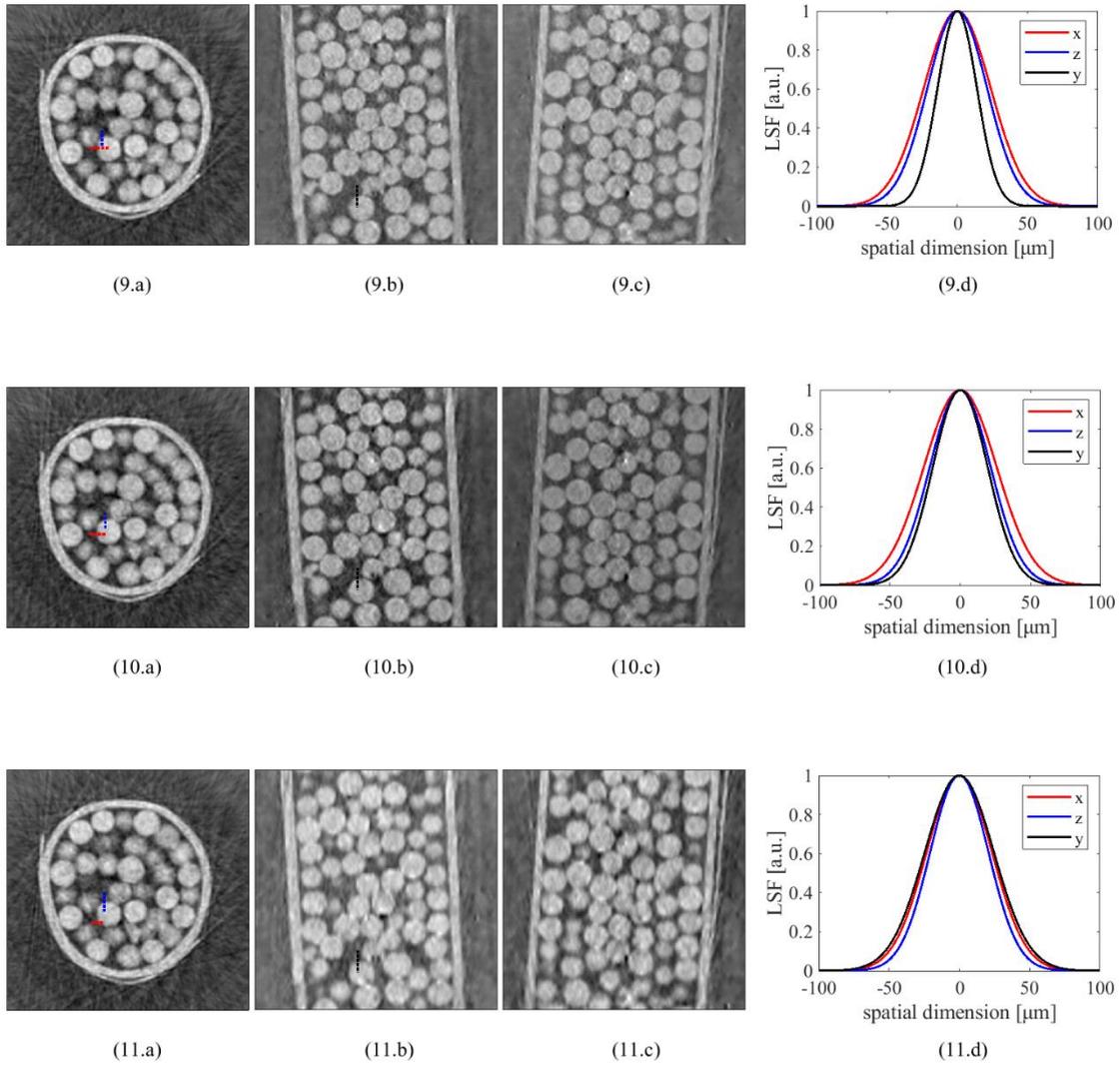

**Fig. SM.** Reconstructed (a) axial, (b) sagittal, and (c) coronal planes of the spheres phantom; (d) the line spread functions along the *x* (red line), *y* (black line), and *z* (blue line) direction, as indicated in the images. Results are shown for sub-sampling the fully sampled (dithered) dataset according to a cycloidal-spiral scheme with (1) $dx = 0$ μm, $dy = 0$ μm (which in fact corresponds to a rotation-only scan), (2) $dx = 10$ μm, $dy = 10$ μm, (3) $dx = 10$ μm, $dy = 20$ μm, (4) $dx = 10$ μm, $dy = 40$ μm, (5) $dx = 20$ μm, $dy = 10$ μm, (6) $dx = 20$ μm, $dy = 20$ μm, (7) $dx = 20$ μm, $dy = 40$ μm, (8) $dx = 30$ μm, $dy = 30$ μm, (9) $dx = 40$ μm, $dy = 10$ μm, (10) $dx = 40$ μm, $dy = 20$ μm, (11) $dx = 40$ μm, $dy = 40$ μm.